\documentclass[12pt,a4paper]{amsart}
\usepackage{amsmath,mathrsfs}
\usepackage{bbm}
\usepackage{graphicx}

\newcommand{\nn}{\nonumber}
\newcommand{\pa}{\partial}
\newcommand{\ben}{\begin{enumerate}}\newcommand{\een}{\end{enumerate}}
\newcommand{\Ref}[1]{(\ref{#1})}

\newcommand{\td}{thermodynamic~}
\newcommand{\elm}{electromagnetic~}

\newcommand{\ga}{\gamma}

\newcommand{\om}{\omega}\newcommand{\Om}{\Omega}

\newcommand{\eq}[2]{\begin{align}\label{#1}#2\end{align}}

\begin{document}
\title{On the entropy of a spherical plasma shell}
\author{M. Bordag}
\address{Universit{\"a}t Leipzig, Institute for Theoretical Physics, Postfach 100920, 04009 Leipzig, Germany}
\email{Michael.Bordag@uni-leipzig.de}
\author{K. Kirsten}
\address{GCAP-CASPER, Department of Mathematics, Baylor University, Waco, TX 76798, USA}
\email{Klaus\textunderscore Kirsten@baylor.edu}
\date{20.9.2018}
\begin{abstract}
Negative entropy was repeatedly observed in the Casimir effect caused by dissipation or geometry. However, it was restricted to subsystems. Recently the question about the entropy for a complete Casimir effect like configuration was raised. In the present paper we consider a spherical plasma shell which can be considered as a (crude) model for a giant carbon molecule (e.g., $C_{60}$). The entropy is free of ultraviolet divergences and its calculation does not need  any regularization. We calculate the entropy numerically and demonstrate unambiguously  the existence of a region where it takes negative values. This region is at small values of temperature and plasma frequency (in units of the inverse radius).
\end{abstract}
\maketitle
%
\section{\label{T0}Introduction}
Entropy is one of the basic notions for physical systems at finite temperature. It raised broad interest over many decades in the past and appears to be a well settled topic at least in classical physics. In quantum physics, new questions about the positiveness of entropy arose  with the observation of negative entropy in the Casimir effect, first in \cite{geye05-72-022111}, later in \cite{brev06-8-236} (for a more recent overview see, e.g.,  the Introduction in \cite{milt17-96-085007}). It must be mentioned that in such systems one is interested in the force between two slabs and only the interaction part of the free energy is considered.
This way, it is the entropy of a subsystem and its negative values alone would not necessarily signal a thermodynamic instability. However, recently in \cite{milt17-96-085007} the entropy of a single system, namely the electromagnetic field in the presence of a sphere carrying a delta function potential, was investigated. Although there is no clear cut result from \cite{milt17-96-085007} {(and a similar paper \cite{liya16-94-085010} on a flat sheet)} due to the divergences appearing in the course of the calculation, the question about the sign of the entropy for a single body was raised.

We mention that negative entropy and negative specific heat were observed repeatedly, starting as early as in \cite{Eddington1926} for stellar systems.  In later work, \cite{thir70-235-339} and \cite{lynd77-181-405}, relations to phase transitions were discussed. More recent interest is related, for example,  to left-handed systems \cite{palm17-471-396}.

In the present paper we investigate the entropy of the electromagnetic field interacting with a plasma sphere, i.e., the entropy of a single body, and show that there is a parameter region where the entropy is negative. Parameters are the temperature $T$, the plasma frequency $\Om$ and the radius $R$ of the sphere. Our result is free of ambiguities since we observe that the entropy does not have any divergences, neither ultraviolet nor infrared, within this model.

The plasma sphere consists of a spherical shell carrying a continuous charged fluid. Its charge is compensated by a homogeneous immobile background of opposite charge, whose entropy we do not account for in this paper.
Para\-meters are the mass $m$,  the charge $e$ and the  density $\rho$ per unit area of its particles. These particles have a non-relativistic dynamics and the usual coupling to the \elm field. The parameters combine into the plasma frequency
\eq{pf}{\Om&=\frac{4\pi e^2\rho}{m c},
}
which is the only parameter entering the equations for the \elm field.
This model was investigated in a series of papers, mainly \cite{BIII} {(where $\mu$ corresponds to $\Om$, eq.(4.11)), and \cite{BV} (where $2q$ corresponds to $\Om$, eq. (2.11))}. It goes back to an earlier model for a two-dimensional electron gas \cite{fett73-81-367}. The model is aimed to describe the $\pi$-electrons in the giant carbon molecule C$_{60}$ and it is called the hydrodynamic model in distinction to the Dirac model well known, e.g., for  graphene. For the C$_{60}$-molecule see, e.g., the book \cite{Dresselhaus1996}, and for more recent interest on this object, e.g., \cite{park00-407-57}. The interaction of the plasma shell with the electromagnetic field is equivalently described by matching conditions the field modes (for TE and TM polarizations)  have to fulfil across the shell. These modes are the degrees of freedom whose thermodynamic properties we investigate in this paper.

For the entropy $S$ we  follow standard textbooks, e.g., \cite{pathria1996}, and use its \td definition as a derivative of the (Helmholtz) free energy $F$,
\eq{1}{S=-\frac{\pa}{\pa T}F.
}
The free energy is related by
\eq{2}{F=-T\ln(Z)
}
to the distribution function $Z$ (denoted by $\mathscr{D}$ in \cite{path98-58-1490}), which is defined as
\eq{3}{Z={\rm Tr}\ e^{-\beta \hat{H}},
}
where $\hat{H}$ is the Hamilton operator of the electromagnetic field together with the matching conditions. It should be mentioned that $\hat{H}$ is hermite, although the spectral problem associated with the TM mode is
somewhat nonstandard due to the occurrence of double poles in the zeta function \cite{bord08-77-085026}. This way, we have a mathematically well defined problem.
The  distribution function $Z$, \Ref{3}, can also be derived from a functional integral in Matsubara representation, see, e.g., \cite{kapu89b}.

The paper is organized as follows. In the next section we introduce the model in detail. In the third and fourth section we calculate the entropy and conclusions are given in the last section. \\
Throughout the paper we use units with $k_{\rm B}=\hbar=c=1$. We use the notation 'TX' in case a formula is valid for   both, 'TE' and 'TM' polarizations.

%
\section{\label{T1}Electromagnetic field and spherical plasma shell}
As said in the introduction, the model consists of the electromagnetic field interacting with a spherical plasma shell of radius $R$. The plasma shell carries a fluid of charged particles with mass $m$ and charge $e$ with density $\rho$. Eliminating the fluid variables from the corresponding equations of motion, the Maxwell equations acquire matching conditions across the shell. This model is described quite detailed in \cite{BIII}, where also older literature is cited. We mention that the polarizations of the \elm field separate into TE- and TM-modes and the mode expansions can be written in the form
\eq{TE1}{
\mathbf{E}^{\rm TE}(t,\mathbf{r})&=
\sum_{{\ell\ge 1\atop  |m|\le l}}\int_0^\infty \frac{d \om}{\pi}
\frac{1}{\sqrt{2\om}}
\left(e^{-i\om t}
g^{\rm TE}_{\ell,m}(r) \mathbf{L} \frac{1}{\sqrt{L^2}}\, Y_{\ell,m}(\vartheta,\varphi)a_{\ell,m}
+c.c. \right),
\nn\\
\mathbf{B}^{\rm TM}(t,\mathbf{r})&=
\sum_{{\ell\ge 1\atop  |m|\le l}}\int_0^\infty \frac{d \om}{\pi}
\frac{1}{\sqrt{2\om}}
\left(e^{-i\om t}
g^{\rm TM}_{\ell,m}(r) \mathbf{L} \frac{1}{\sqrt{L^2}}\, Y_{\ell,m}(\vartheta,\varphi)b_{\ell,m}
+c.c. \right),
}
where the second line follows by duality and $a_{\ell,m}$  and $b_{\ell,m}$  are free coefficients. We mention that the sums start from $\ell=1$ since the s-wave is absent in the considered model, in distinction to, e.g., the scalar field. The functions $g^{\rm TX}_{\ell}(r)$ are the scattering solutions,
\eq{scs}{  g^{\rm TX}_{\ell,m}(r) &
=j_\ell(\om r) \Theta(R-r)               +\frac12
\left(f^{\rm TX}_\ell(\om)h^{(2)}_\ell(\om r)
+{f^{\rm TX}_\ell(\om)}^*h^{(1)}_\ell(\om r)\right)\Theta(r-R),
}
expressed in terms of the spherical Bessel functions and of the Jost functions $f^{\rm TX}_\ell(\om)$.
The matching conditions on the \elm field imply the following conditions on the mode functions,
\eq{mc}{g^{\rm TE}_{\ell,m}(r)_{\mid_{r=R+0}}& = g^{\rm TE}_{\ell,m}(r)_{\mid_{r=R-0}}, &\left(rg^{\rm TE}_{\ell,m}(r)\right)'_{\mid_{r=R+0}}-
\left(rg^{\rm TE}_{\ell,m}(r)\right)'_{\mid_{r=R-0}}&= \Om Rg^{\rm TE}_{\ell,m}(R),
\nn\\
\left(rg^{\rm TM}_{\ell,m}(r)\right)'_{\mid_{r=R+0}}&=
\left(rg^{\rm TM}_{\ell,m}(r)\right)'_{\mid_{r=R-0}},
& g^{\rm TM}_{\ell,m}(r)_{\mid_{r=R+0}} - g^{\rm TM}_{\ell,m}(r)_{\mid_{r=R-0}}
&=-\frac{\Om}{\om^2R} \left(rg^{\rm TM}_{\ell,m}(r)\right)'_{\mid_{r=R}},
}
where $\Om$ is the plasma frequency.
For the TE mode these conditions are equivalent to a delta function potential $\Om\delta(r-R)$ in the wave equation for a scalar field (in that case one has to include the s-wave ($\ell=0$) contribution). For the TM mode this condition can be related to a delta' potential with frequency dependent coupling -- a relation which we do not use in the present paper.

Inserting \Ref{scs} into \Ref{mc} the expressions
\eq{jfs}{   f^{\rm TE}_{\ell}(\om) &=
            1+\frac{\Om}{\om} \ \hat{j}_\ell(\om R) \hat{h}^{+}_\ell(\om R),
\\\nn     f^{\rm TM}_{\ell}(\om) &=
            1+\frac{\Om}{\om} \ \hat{j}'_\ell(\om R) \hat{h}^{+'}_\ell(\om R)
}
follow representing the Jost functions in terms of the Riccati-Bessel functions, $\hat{ j}_\ell (z)=\sqrt{\frac{\pi z}{2}}J_{l+\frac12}(z)$,
 $\hat{ h}^+_\ell (z)=i \sqrt{\frac{\pi z}{2}}H^{(1)}_{l+\frac12}(z)$. Defining the phase shifts as usual by
\eq{ps}{ \frac{f_\ell(\om)}{f_\ell(\om)^*}=e^{-2i\delta_\ell(\om)}
}
one arrives at
\eq{tan}{\tan \delta_\ell^{\rm TE}=
\frac{-\frac{\Om}{\om}\left(\hat{j}_\ell(\om R)\right)^2}
{1-\frac{\Om}{\om} \, \hat{j}_\ell(\om R)\hat{y}_\ell(\om R)},
 \ \ \tan \delta_\ell^{\rm TM}=
\frac{-\frac{\Om}{\om}\left(\hat{j}'_\ell(\om R)\right)^2}
{1-\frac{\Om}{\om} \, \hat{j}'_\ell(\om R)\hat{y}'_\ell(\om R)} ,
}
with $\hat{y}_\ell (z) = \sqrt{ \frac{\pi z} 2 } N_{\ell + \frac 1 2} (z)$ (see \cite{BIII}, eqs. (4.11) and (4.12)). For completeness we also note the Jost functions of imaginary argument,
\eq{jim}{ f^{\rm TE}_{\ell}(i\xi)=1+\frac{\Om}{\xi}\,s_\ell(\xi R)e_\ell(\xi R),
    \ \ f^{\rm TM}_{\ell}(i\xi)=1-\frac{\Om}{\xi}\,s'_\ell(\xi R)e'_\ell(\xi R),
}
expressed in terms of the modified Riccati-Bessel functions.

From \Ref{tan}, the phase shifts can be expressed as
\eq{dTX}{ \delta^{\rm TE}_\ell(\om)
            =\arctan \frac{-\frac{\Om}{\om} \, \hat{j}_\ell(\om R)^2}
        {1-\frac{\Om}{\om} \, \hat{j}_\ell(\om R)\hat{y}_\ell(\om R)}
,\ \     \delta^{\rm TM}_\ell(\om)=
        -\frac{\pi}{2}+\arctan \frac{1-\frac{\Om}{\om} \, \hat{j}'_\ell(\om R)\hat{y}'_\ell(\om R)}
         {\frac{\Om}{\om} \, \hat{j}'_\ell(\om R)^2}
}
with $\delta_\ell^{\rm TX}(\om) \raisebox{-4pt}{$\to \atop \om\to\infty$}       0$.
The behavior for small $\om$ is dominated by the s-wave ($\ell=1$) for the TE-polarization and by a constant for the TM-polarization,
\eq{low}{ \delta_\ell^{\rm TE}(\om) \raisebox{-4pt}{$\sim \atop \om\to0$}
            -\frac{\Om R^4}{3 (3+\Om R)}\om^3 \delta_{\ell,1}+\dots, \ \
\delta_\ell^{\rm TM}(\om) \raisebox{-4pt}{$\sim \atop \om\to0$}  -\pi+\frac{2R^3}{3 }\om^3\delta_{\ell,1}+\dots.
}
In the expansion of $\delta_1^{\rm TM}(\om) $ the dependence on $\Om$ appears only in higher orders { (see the remark on the Klauder phenomenon below in Sect. 4)}. For  a discussion of the relation to  Levinson's theorem see footnote 15 in \cite{BIII}.

The sum of phase shifts is defined by
\eq{sps}{\delta_{\rm TX}(\om)
        =\sum_{\ell=1}^\infty (2\ell+1)\delta_\ell^{\rm TX}(\om).
}
For the TE modes, the convergence of the sum for $\ell\to\infty$ is very fast,
\eq{dlinf}{ \delta_\ell^{\rm TX}
\raisebox{-4pt}{$\sim \atop \ell\to\infty$}             \ell^{-2\ell},
}
 as can be seen from the asymptotic expansions of the Bessel functions for large index.
This way, for the TE polarization, $\delta_{\rm TE}(\om)$ is a well behaved function of its argument. In contrast, for the TM-polarization the sum of phase shifts does not exist because of their behavior for $\om\to0$. However, below we will need only the derivatives of the sums over phase shifts,
\eq{ds}{ \delta'_{\rm TX}(\om)\equiv\frac{\pa}{\pa\om}\delta_{\rm TX}(\om)
    =\sum_{\ell=1}^\infty(2\ell+1)\frac{\pa}{\pa\om}\delta_\ell^{\rm TX}(\om),
}
and the mentioned problem does not occur. The convergence for $\ell\to\infty$ is as fast as in \Ref{sps}. The expansions for $\om\to 0$, following from \Ref{low}, read,
\eq{lows}{ \delta'_{\rm TE}(\om) \raisebox{-4pt}{$\sim \atop \om\to0$}
                            -\, 3 \cdot \frac{\Om R^4}{3+\Om R}\om^2+\dots,
\ \ \delta'_{\rm TM}(\om) \raisebox{-4pt}{$\sim \atop \om\to0$}   3\cdot  2R^3\om^2+\dots.
}
The derivatives of the sum of phase shift functions are shown in Fig. \ref{fig1}. As can be seen, the function $\delta'_{\rm TE}(\om)$ is a quite smooth function
whereas $\delta'_{\rm TM}(\om)$ has (finite) peaks growing in size and narrowing with increasing $\om$.
These peaks are located at or very close to $\om_c$, which is defined to be the solution of the equation
\eq{omc}{1-\frac \Omega \omega \hat j _\ell ' (\omega R ) \hat y _\ell ' (\omega R) =0 ,
}
which is where the argument of the $\arctan$ in (\ref{dTX}) goes through zero.
It is interesting to mention that the $\om_c$ in the limit of small $\om$ turn into the plasmon normal modes discussed in \cite{bart91-95-1512} (eq.~(2.6)).
For large $\ell$ these are located at  $\sqrt{\frac{\Om }{2 R}\left(\ell+\frac12\right)}$ as can be derived from Debye's expansion of the Bessel functions.
\setlength{\unitlength}{1cm}\begin{figure}[h]
\begin{picture}(18,6)
\put(0,0){\includegraphics[width=8cm]{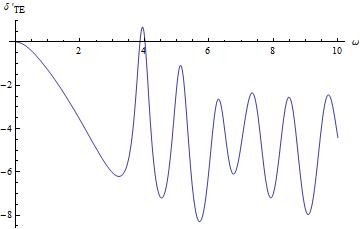}}\put(9,0){\includegraphics[width=8cm]{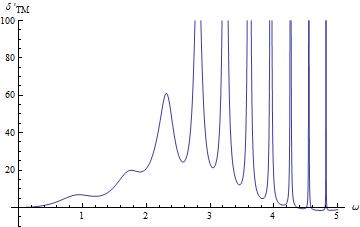}}
\end{picture}
 \caption{The derivatives of the sums of phase shifts for $\Om=5$ and $R=1$.}\label{fig1}  \end{figure}

%
\section{\label{T2}Free energy and entropy}
The free energy, as defined in \Ref{2} can be related to the spectral problem associated with the \elm field discussed in the preceding section by
\eq{2.1}{   F=\sum_J\left(\frac{\om_J}{2}+T\ln\left(1-e^{-\om_J/T}\right)\right),
}
where $\om_J$ are the eigenfrequencies of the modes of the \elm field subject to the matching conditions \Ref{mc}. Such representation for the free energy is the generalizations of the corresponding formula for a single oscillator to a set characterized by the $\om_J$. It was used for example for a conducting surface, eq. (5.1) in  \cite{bali78-112-165}. In \Ref{2.1} the summation index $J$ assumes a discrete spectrum, whereas in our model it will be continuous.
Let us briefly describe the well known procedure how to use representation \Ref{2.1} in that case. One has to put the system into a large sphere of radius $\tilde{R}$ and to impose Dirichlet (or other) boundary conditions on that sphere,
\eq{2.2}{ g_{\ell,m}^{\rm TX}(\tilde{R})=0.
}
The solutions $\om=\om_J$ of this equation are discrete eigenfrequencies. Then one transforms the sum over $J$ into a contour integral,
\eq{2.3}{F=
\sum_{TX=TE,TM} \sum_{{\ell\ge 1\atop  |m|\le l}}\int_\ga \frac{d \om}{2\pi i}
\left(\frac{\mu^{2s}\om^{1-2s}}{2}+T\ln\left(1-e^{-\om/T}\right)\right)
\frac{\pa}{\pa\om} \ln g_{\ell,m}^{\rm TX}(\tilde{R}),
}
where the path $\ga$ encircles the real zeros of $g_{\ell,m}^{\rm TX}(\tilde{R})$.
In \Ref{2.3} we also introduced the zeta functional regularization with parameter $s>\frac32$ for the ultraviolet divergencies and the arbitrary parameter $\mu$ having the dimension of a mass. The next step is to remove the large sphere by tending $\tilde{R}\to\infty$. This procedure is also well known and described in detail, e.g., in \cite{bord96-53-5753} or in the book \cite{kirs01b}. Dropping the empty space contribution one ends up with the formula
\eq{2.4}{F=E_0(s)+\Delta_TF,
}
where
\eq{2.5}{E_0(s)=
  \frac{\mu^{2s}}{2} \int_0^\infty\frac{d\om}{\pi}
\om^{1-2s} \delta'(\om)
}
is the vacuum energy
and
\eq{2.6}{ \Delta_TF= T\int_0^\infty\frac{d\om}{\pi}
\ln\left(1-e^{-\om/T}\right) \delta'(\om)
}
is the temperature dependent part of the free energy. In the temperature dependent part
we could put $s=0$ since there are no ultraviolet divergencies due to the decrease of the integrand. In \Ref{2.5} and \Ref{2.6} we also introduced the notation
\eq{2.7}{\delta'(\om)=\delta'_{\rm TE}(\om)+\delta'_{\rm TM}(\om),
}
where the $\delta'_{\rm TX}(\om)$ were defined in \Ref{ds}.

We mention that it is frequently possible to integrate by parts,
\eq{2.8}{
E_0(s)= -\frac{\mu^{2s}}{2}(1-2s)  \int_0^\infty\frac{d\om}{\pi}
\om^{-2s} \delta(\om), \ \
\Delta_TF = -\int_0^\infty\frac{d\om}{\pi}\frac{1}{e^{\om/T}-1}\delta(\om).
}
It depends on the properties of the phase shifts $\delta(\om)$ whether this is possible or whether boundary terms appear. We will not use this representation in the present paper. Also we mention that it is possible to make a Wick rotation. For that one has to go back to phase shifts for the individual orbital momenta \Ref{ds} and further to the Jost function \Ref{jim}. One comes to
\eq{2.9}{E_0(s)=-\frac{\cos(\pi s)}{2\pi}\mu^{2s} \sum_{\ell=1}^\infty (2\ell+1)
\int_0^\infty d\om\  \om^{1-2s}\frac \partial {\partial \omega} \ln f_\ell(i\omega)
}
and the usual Matsubara sum for $\Delta_TF$.\\
Let us stress that in the above formulas it is assumed that there are no bound states in the spectrum. These would give additional contributions.
It is well known that in our model, \elm field and matching condition \Ref{mc},  there are no bound states (we consider only $\Om>0$, see \Ref{pf}). This may be seen from the Jost functions \Ref{jim} at imaginary frequency which have no poles. { The vacuum energy \Ref{2.9} was calculated in \cite{bord08-77-085026}.  }

In the present paper we are interested in the entropy $S$ as given by eq. \Ref{1}. Because of the derivative, the vacuum energy does not contribute and
we get from \Ref{1} and \Ref{2.6}
\eq{2.10}{ S
    =\int_0^\infty\frac{d\om}{\pi}g\left(\frac{\om}{T}\right)\delta'(\om)
}
with
\eq{2.11}{g(x)=\frac{x}{e^x-1}-\ln\left(1-e^{-x}\right).
}
Due to the decrease of the function $g \left(\frac{\om}{T}\right)$ in the integrand, there are no ultraviolet divergencies in the entropy. Nevertheless $\Delta_TF$, \Ref{2.6}, and $S$, \Ref{2.10}, cannot be considered as physical. The problem is in their behavior at high temperature. {  As known (see, for example, eq. (5.51) in \cite{BKMM}),} for $T\to\infty$ we have
\eq{2.12}{   F=
    -\frac{\zeta(3)a_{\frac12}}{4\pi^{3/2}} \frac{(k_{\rm B}T)^3}{(\hbar c)^2}
    -\frac{a_1}{24}  \frac{(k_{\rm B}T)^2}{ \hbar c}
    {-\frac{a_{\frac32}}{(4\pi)^{3/2}}k_{\rm B}T\ln(k_{\rm B}T)} +O(T),
}
where we restored for the moment the dependence on $k_{\rm B}$, $\hbar$ and $c$.
Here, the $a_k$ are the heat kernel coefficients. We wrote done the dependence on $\hbar$ and $c$ explicitly (usually it is included in the coefficients).  For the considered model the relevant ones were calculated in \cite{bord08-77-085026} and read
\eq{hkk}{
        a_{\frac12}^{\rm TE} &=0,  &a_{1}^{\rm TE} &= -4\pi\Om R^2,
      & {a_{\frac32}^{\rm TE}}&  {= \pi^{3/2}\Om^2 R^2},
\nn\\  a_{\frac12}^{\rm TM} &=8\pi^{3/2}R^2,  &a_{1}^{\rm TM} &= -\frac{4\pi}{3}\Om R^2,
      & {a_{\frac32}^{\rm TM}}&{= -\frac{10}{3}\pi^{3/2}}.
}
{  We mention that the coefficients with $k\ge1$, being divided by $4\pi R^2$, turn for $R\to\infty$ into that of a flat sheet. The coefficients $a_\frac12$ for both polarizations are nonzero for the flat sheet (see eqs. (4.14), (4.28) and (4.29) in \cite{bord05-38-11027} where $q$ corresponds to $\Om/2$). }

Inserting \Ref{hkk} into eq. \Ref{2.12} we get for the considered model
\eq{2.12a}{   F=
    -2\zeta(3) R^2 \frac{(k_{\rm B}T)^3}{(\hbar c)^2}
    +\frac{2\pi\Om R^2}{9}\,\,  \frac{(k_{\rm B}T)^2}{ \hbar c}
    -\frac18\left(\Om^2R^2-\frac{10}{3}\right) k_{\rm B}T\ln(k_{\rm B}T) +O(T).
}
The next terms in \Ref{2.12} and \Ref{2.12a} { are  a term linear  in $T$,  a constant, and terms involving inverse powers of $T$}.
The $T\ln(T)$-term and the term linear in $T$ form the so called classical limit since these do not involve $\hbar$. All higher orders of the expansion can be expressed in terms of the heat kernel coefficients, see eq. (5.51) in \cite{BKMM}, and are proportional to positive powers of $\hbar$.

As can be seen the displayed contributions have $\hbar$ in their denominators. Therefore these must be considered as non physical and must be subtracted
in analogy to the ultraviolet renormalization in the sense of a finite renormalization. An equivalent reason for that is the classical limit of the entropy. { As known, see. e.g., eq. (1.5) in \cite{wehr78-50-221}, the entropy of any quantum system does not exceed its classical limit. But in the classical limit the growth of the free energy with temperature does not  exceed the first power (up to logarithmic contributions).}

Defining
\eq{2.13}{F^{\rm high}=
-2\zeta(3)R^2T^3+\frac{2\pi\Om}{9}R^2T^2
}
and, using \Ref{1},  accordingly
\eq{2.14}{S^{\rm high}=6\zeta(3)R^2T^2-\frac{4\pi\Om}{9}R^2T ,
}
we introduce
\eq{2.15}{ \Delta_T F^{\rm subtr}=\Delta_T F-F^{\rm high}
}
and
\eq{2.16}{  S^{\rm subtr}=S-S^{\rm high}
}
as the corresponding physical quantities. For a more detailed discussion of this procedure see chapt. 5 in \cite{BKMM} and literature cited therein, {  especially \cite{geye08-57-823}}.

Out next point is to consider the low temperature behavior. We consider the temperature dependent part of the free energy as given by eq. \Ref{2.6}. After the substitution $\om\to \om T$ we use the expansion \Ref{lows} and get with \Ref{2.7}
\eq{2.16a}{\Delta_T F&=T^2\int_0^\infty \frac{d\om}{\pi} \ln\left(1-e^{-\om}\right)
    \left(\left( 2-\frac{\Om R}{3+\Om R}\right) R^3\om^2 T^2+\dots\right)
\nn\\&=-\frac{\pi^3}{15}\,\,\frac{6+\Om R}{3+\Om R}R^3T^4+\dots
}
and, using \Ref{1},
\eq{2.17}{S &=\frac{4\pi^3}{15}\,\,\frac{6+\Om R}{3+\Om R}R^3T^3+\dots\,.
}
These contributions are subleading in $\Delta_T F^{\rm subtr}$ and $S^{\rm subtr}$ whose expansions for $T\to0$ read
\eq{2.18}{  \Delta_T F^{\rm subtr}&=
-\frac{2\pi\Om R^2}{9}T^2+2\zeta(3)R^2T^3-\frac{\pi^3}{15}\,\,\frac{6+\Om R}{3+\Om R}R^3T^4+\dots\,,
\nn \\
S^{\rm subtr}&= \frac{4\pi\Om R^2}{9}T-6\zeta(3)R^2T^2+\frac{4\pi^3}{15}\,\,\frac{6+\Om R}{3+\Om R}R^3T^3+\dots\,.
}
The expansions can be continued to higher orders by including higher orders of the expansions \Ref{lows}. { We mention that these expansions hold if $T$ is smaller than all other dimensional quantities involved.

The limit for $T\to\infty$ is
\eq{2.19}{\Delta_T F^{\rm subtr}&=-\frac18\left(\Om^2R^2-\frac{10}{3}\right) T\ln(T)+\dots,
\nn\\   S^{\rm subtr}&=\frac18\left(\Om^2R^2-\frac{10}{3}\right) \ln(T)+\dots\,.
}
As can be seen, for $\Om R<\sqrt{10/3}$, $S^{\rm subtr}$ is negative.}

With formulas \Ref{2.10}, \Ref{2.14} and \Ref{2.16}, together with the expressions for the phase shifts in terms of sums of Bessel functions, we can calculate the entropy $S^{\rm subtr}$ numerically. Its behavior for low temperature is given by eq. \Ref{2.18}, for high temperature by the classic limit {  \Ref{2.18}}.  Examples will be presented in the next section.

%
\section{\label{T4}Numerical evaluation of the entropy}
The main purpose of this section is to show that the entropy for the spherical plasma shell can be negative.
In this section we will set $R=1$ as the dependence on $R$ can be recovered by purely dimensional reasoning. In order to observe that
a negative entropy is indeed possible let us restate eq. (\ref{2.14}) and (\ref{2.16}) using numerical values,
\begin{eqnarray}
S^{high} &=& 7.21234\,\, T^2 - 1.39629 \,\,\Omega T, \nonumber\\
S^{subtr} &=& S - 7.21234\,\, T^2 + 1.39629 \,\,\Omega T . \label{sneg1}
\end{eqnarray}
As the plasma frequency $\Omega \to 0$, the configuration obtained is free space and the associated value for
the entropy $S$ is $S^{free} =0$. So if we had $S\to 0$ as $\Omega \to 0$, then clearly $S^{subtr} < 0$ is possible.
This is so because for $\Omega $ small enough the second term in $S^{subtr}$ above would become the leading contribution and a negative entropy would follow.

However, this limiting behavior is by no means guaranteed as is clear from the heat kernel coefficients (\ref{hkk}), which
clearly show that as $\Omega \to 0$ one does not recover the heat kernel of free space! This phenomenon is referred to as the Klauder
phenomenon \cite{klau73-11-341,klau73-47-523}, which roughly states that sufficiently singular perturbations
{\it cannot} be turned off to restore the unperturbed free situation \cite{simo73-14-295,defa74-15-1071}.
However, as we will observe numerically, $S$ gets smaller for smaller $\Omega$, which will render a negative entropy possible.

The main numerical complication in computing the entropy is the pathological behavior of $[\delta_\ell ^{TM} (\omega ) ]'$. It
has a very sharp spike at the frequency $\omega =\omega_c$, defined in eq. \Ref{omc}.
The height of that spike is increasing rapidly with $\ell$, whereas the width of the spike is decreasing rapidly with $\ell$.
For $\Omega =5$ the above description is substantiated in Figure \ref{fig1}. For smaller values of $\Omega$ the behavior is so extreme
that it cannot be meaningfully represented in a graph. Instead, for $\Omega = 0.05$ and $T=0.0105$, in Table \ref{tab1}, we have summarized essential features.
As is clearly seen from these numbers, the combined effect of increasing height and decreasing width is that the contribution of the spike to the entropy is decreasing with $\ell$ and in the relevant
range of parameters only very few angular momenta $\ell$ have to be summed to get very accurate values for the entropy.
\begin{table}
\begin{tabular}{|c|c|c|c|c|c|} \hline
$\ell$ & $\omega_c$ & $[\delta_\ell ^{TM} (\omega_c )]'$ & $|\omega_c - \omega_h|$ & $g (\omega_c /T) [\delta_\ell^{TM} (\omega_c)]'$ & $|\omega_c - \omega_h | g (\omega_c /T) [\delta_\ell ^{TM} (\omega_c )]'$\\ \hline
$1$ & $0.18052691$ & $2860.2426$  & $0.00035129$ & $ 0.00177610$ & $6.23928\cdot 10^{-7}$   \\  \hline
$2$ & $ 0.24406326$ & $287976$  & $3.47228\cdot 10^{-6}$ & $0.00056127$ & $1.94888\cdot 10^{-9}$ \\ \hline
$3$ & $ 0.29214134$ & $4.51670\cdot 10^7$ & $2.21400\cdot 10^{-8}$ & $0.00107446$ & $2.37887\cdot 10^{-11}$ \\ \hline
$4$ & $0.33281602$ & $9.65673\cdot 10^9$ & $1.03555\cdot 10^{-10}$ & $0.00541509$ & $5.60758 \cdot 10^{-13}$ \\ \hline
$5$ & $0.36882255$ & $2.62036\cdot 10^{12}$ & $3.81627\cdot 10^{-13}$ & $0.05262237$ & $2.00821 \cdot 10^{-14} $ \\ \hline
$6$ & $0.40151099$ & $8.62814 \cdot 10^{14}$ & $1.15900 \cdot 10^{-15}$ & $0.83673547$ & $9.69775 \cdot 10^{-16}$ \\ \hline
$7$ & $ 0.43167475$ & $3.34282 \cdot 10^{17}$ & $ 2.99149 \cdot 10^{-18}$ & $19.672272$ & $5.88493 \cdot 10^{-17}$ \\ \hline
\end{tabular}\\[.2cm]
\caption{In this table, $\omega_c$ denotes the location of the peak of $[\delta_\ell^{TM} (\omega) ]'$, $[\delta_\ell^{TM} (\omega_c) ]'$ is the height of the peak, $\omega_h$ is a frequency value
such that $[\delta_\ell^{TM} (\omega_h) ]' = [\delta_\ell^{TM} (\omega_c) ]'/2$. The entry $|\omega_c - \omega_h|$ is an estimate for the width of the peak,
and consequentially the last column gives an estimate for the contribution of the peak to the entropy (\ref{2.10}).  } \label{tab1}
\end{table}

In order to get a rough idea about where to search for negative entropy, note that for sufficiently small $\Omega$, from (\ref{sneg1}) we want to search
the region of parameters where
$$-0.721234 \, T^2 + 1.39629 \, \Omega T \sim 0,$$
so more or less for $\Omega \sim 5T$. For $\Omega$ sufficiently small, this is indeed exactly the relevant range as seen in Figure \ref{fig5}.
If $\Omega$ is too large, $S$ compensates the negative contribution from $S^{high}$ in $S^{subtr}$ and a positive entropy results; see Figure \ref{fig5}.
However, it is clearly seen that for small enough $\Omega$ a temperature range with corresponding negative entropy exists.

{  The above numerical calculation is valid for small $\Om R$ and for small $T$. For large $T$ more sophisticated numerical methods are necessary, especially due to the mentioned peaks in the TM case. Thus for large $T$ we are left with the analytical result \Ref{2.19}. }
\setlength{\unitlength}{1cm}\begin{figure}[h]
\begin{center}
\begin{picture}(18,11.5) \label{fig5}
\put(9,5.5){\includegraphics[width=8cm]{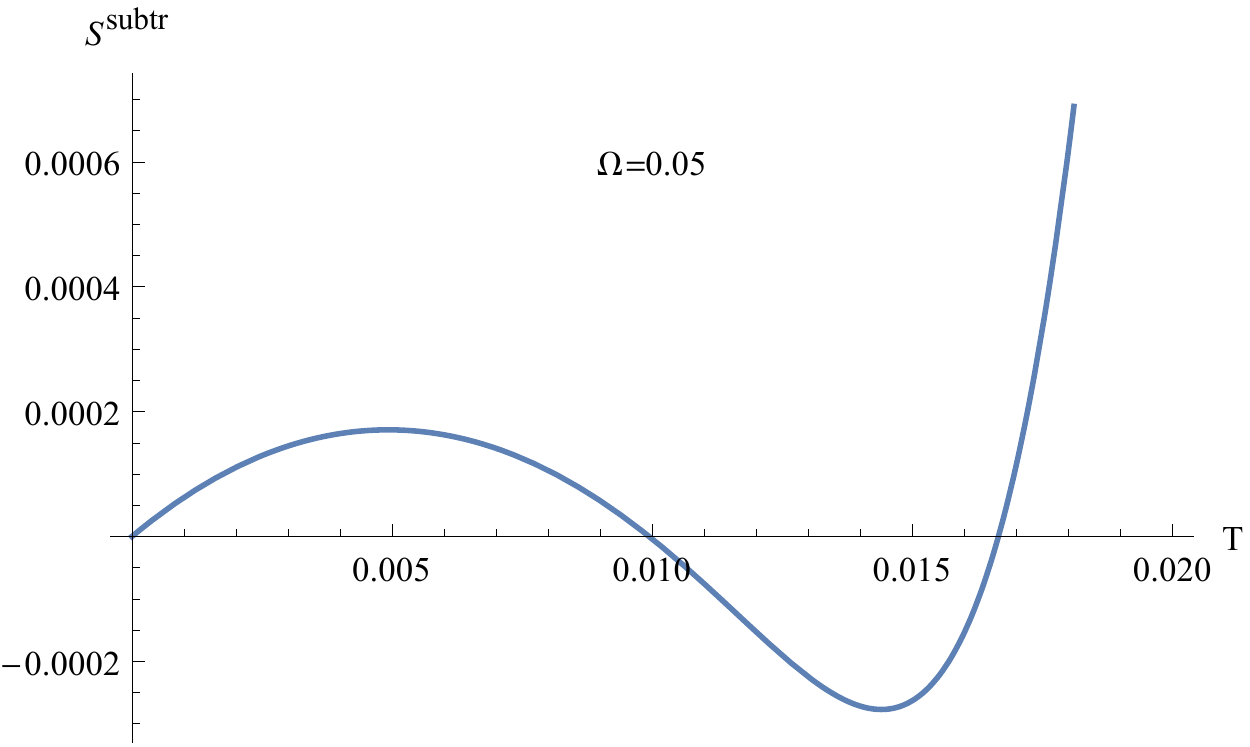}}
\put(0,6.5){\includegraphics[width=8cm]{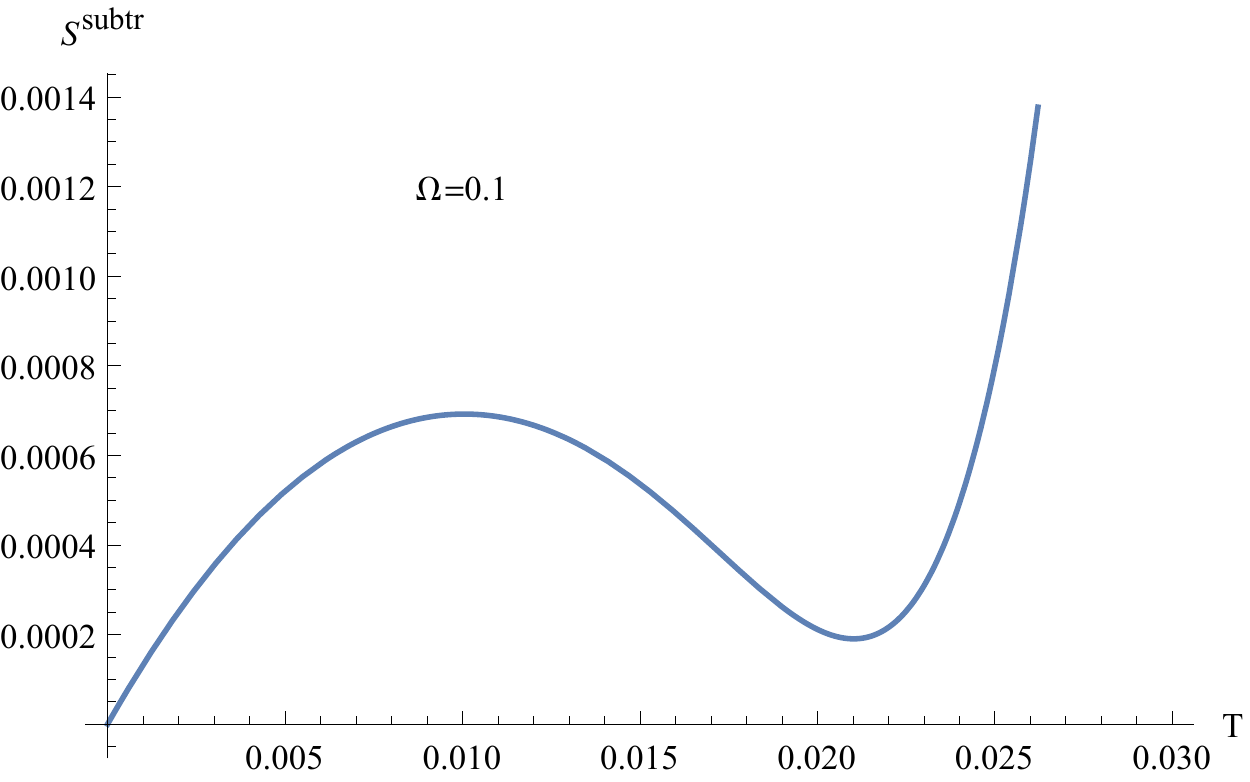}}
\put(0,0.5){\includegraphics[width=8cm]{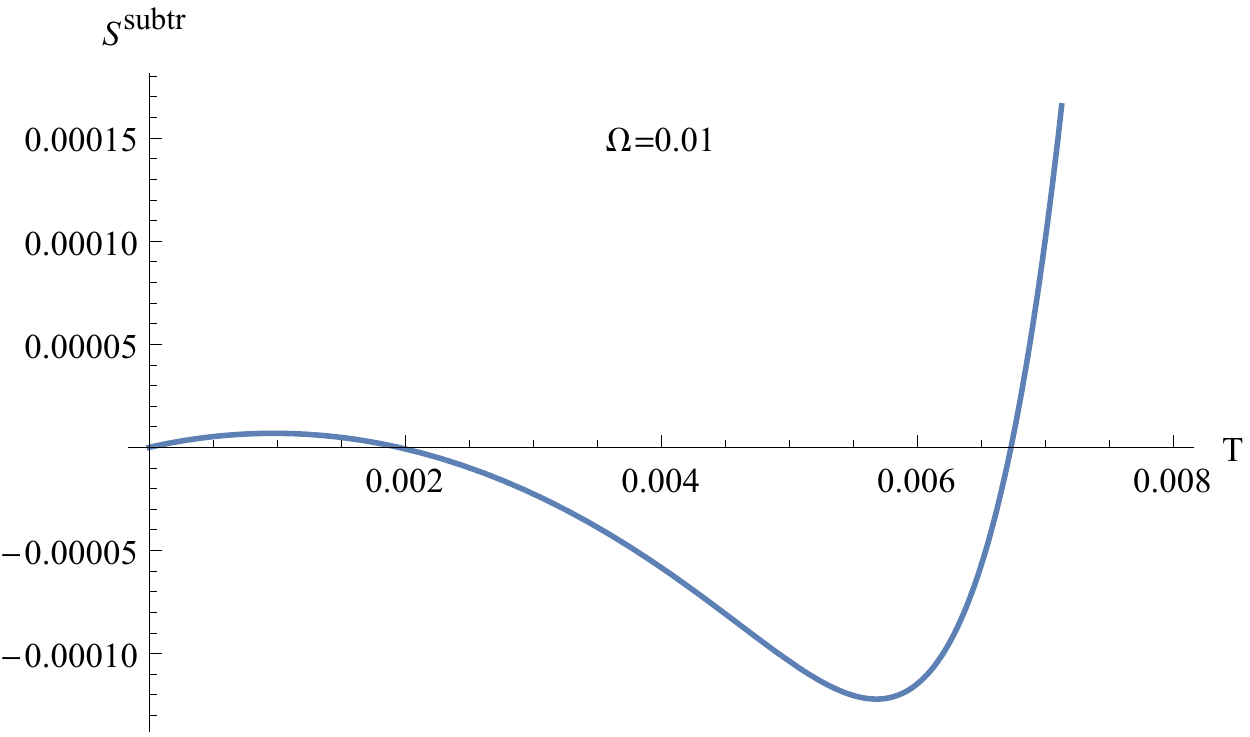}}
\put(9,0.5){\includegraphics[width=8cm]{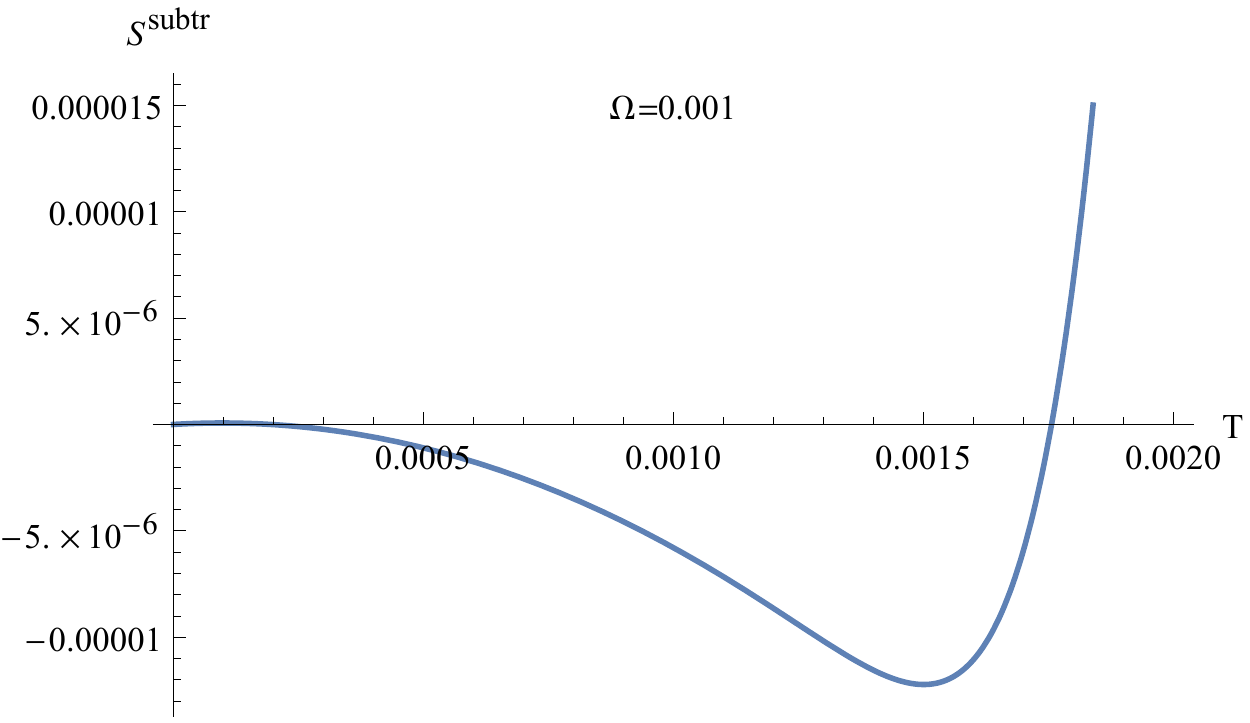}}
\end{picture}
\end{center}
\caption{Total entropy  as a function of $T$ at $R=1$. From top left, $\Om=0.1$, $\Om=0.05$, $\Om=0.01$ and $\Om=0.001$, demonstrating the appearance of the region with negative entropy. }   \end{figure}
%

\section{\label{T5}Conclusions}
In the forgoing sections we considered a model consisting of the \elm field interacting with a dispersive spherical plasma shell. The free energy and the entropy are given by eqs. \Ref{2.4}-\Ref{2.6} and \Ref{2.10} in terms of the derivative of the summed scattering phase shifts,
\eq{5.1}{S=\int_0^\infty\frac{d\om}{\pi}
\left(\frac{\om/T}{e^{\om/T}-1}-\ln\left(1-e^{-\om/T}\right)\right)
\delta'(\om).
}
An equivalent representation can be given after Wick rotation in terms of Matsubara frequencies. Also, such representation can be obtained within the considered model if coupling the plasma degrees of freedom to a heat bath as recently discussed in \cite{bord17-96-062504} for the interaction of polarizable bodies. An application of this dissipative approach to the plasma sphere would be an interesting generalization. Also we mention that similar formulas can be obtained within an even more general approach like that resulting in the remarkable formula in \cite{ford85-55-2273}.

It is interesting to mention that in \cite{bord17-96-062504} it was seen that formulas like \Ref{2.4}-\Ref{2.6} give the complete free energy, i.e., including that of the polarizable fluid and the heat bath. It was also shown that this is the complete free energy of the system if
the dissipation is switched off. As mentioned in the introduction, the entropy of the neutralizing background was not accounted for.

The free energy $F$ consists of the vacuum energy $E_0$ and the temperature dependent part $\Delta_TF$. The vacuum energy has the known ultraviolet divergences.  These can be subtracted within a renormalization procedure. A physical interpretation of this procedure, fixing the known arbitrariness uniquely, i.e., including the freedom of a finite renormalization,  was given in \cite{bord08-77-085026} for the  model considered in the present paper by demanding that in the limit $\Om\to\infty$, i.e., for large plasma frequency, Boyer's result \cite{boye68-174-1764} on ideally reflecting sphere must (and could) be reproduced.

In contrast to the vacuum energy, the temperature dependent part $\Delta_T{ F}$ of the free energy, and with it the entropy $S$, are free of ultraviolet divergencies. It is to be mentioned that this circumstance makes the use of any regularization unnecessary. However, $\Delta_T{ F}$ and $S$ need for a finite renormalization nevertheless. As discussed in Sect. 3, there are contributions involving negative powers of Planck's constant (see eq. \Ref{2.12}) which must be subtracted as being unphysical.
These subtractions can be considered as a finite renormalization which must be done in addition to the ultraviolet renormalization of the vacuum energy. Within a counterterm interpretation of the renormalization, this procedure implies temperature dependent counter\-terms. This way, also the entropy, although itself being free from ultraviolet divergences, needs a renormalization. In addition we mention that the high temperature expansion of the free energy cannot have higher powers than the linear one (and $T\ln T$) in \Ref{2.12}, which is the classical limit, and all other contributions must be proportional to positive powers of $\hbar$ as the contributions $O\left(\frac{1}{T}\right)$ in \Ref{2.12} do (see eq. (5.53) in \cite{BKMM}). We denoted the contributions from $a_{\frac 1 2}$ and $a_1$ in \Ref{2.12}, i.e., the contributions with negative powers of $\hbar$,  by $S^{\rm high}$, \Ref{2.14}, and considered  $S^{\rm subtr}$, \Ref{2.16}, as the physical entropy of the considered model.

In section 4 we investigated the entropy $S^{\rm subtr}$
numerically and showed that there is a region in the space of parameters $T$ and $\Om$ (in units of the radius of the sphere), where this entropy takes negative values.
This region is quite small and located at small values of $T$ and $\Om$. These results are shown in a sequence of figures, Figs. 2 to 5, from larger values of $\Om$, where $S$ is completely positive, to smaller values, where it has a region with negative values.  We mention that the renormalized vacuum energy also has a small parameter region, where it takes negative values \cite{bord08-77-085026}.
{In Sect. 3 we have seen that for high temperature  the entropy will take negative values for $\Om R<\sqrt{10/3}$.}
These are the main results of our paper.

{ It is interesting to discuss the relation of our results with those for the ideally conducting sphere obtained in \cite{bali78-112-165}. The temperature dependence of the free energy \Ref{2.6} should directly turn into that of a conducting sphere, consisting of the sum over the discrete eigenfrequencies inside and an integral over the frequencies in the outside region where the phase shifts are obtained as the formal limit $\Om\to\infty$ in \Ref{dTX}. The leading behavior for $T\to\infty$ can be obtained from \Ref{2.12} with the heat kernel coefficients of a conducting sphere. These are well known, see for example eq. (9.132) in \cite{BKMM}. Taking TE and TM as well as inside and outside regions together, one has $a_\frac12=a_1=0$ and $a_\frac32=2\pi^{3/2}$. Thus there are no subtraction terms (for taking inside or outside or one of the polarisations alone there would be subtraction terms) and the leading order for $T\to\infty$ follows from \Ref{2.12} to be
\eq{5.2}{ { F}_{\rm cond. sphere}=-\frac14 T\ln(T)+\dots\,,
}
in agreement with \cite{bali78-112-165}, eq. (8.39).

Clearly the limit $\Om\to\infty$ in \Ref{2.19} does not reproduce the free energy \Ref{5.2}  of the conducting sphere. The reason is in the heat kernel coefficients \Ref{hkk}, which for $\Om\to\infty$ do not turn into those of the conducting sphere. Mathematically, one has a singular potential like a delta function and noncommuting limits. Physically, for the vacuum one performs the renormalization in a way removing all contributions from the coefficients $a_k$ with $k\le2$ leaving a renormalized result which for $\Om\to\infty$ turns into that of the conducting sphere as done in \cite{bord08-77-085026}. For the free energy, i.e., for the temperature dependent part, this procedure works with the contributions proportional to $T^2$ and to $T^3$, but not for the $T\ln(T)$-contribution.  We keep this observation as an open question. We would like to mention that in the approach taken in \cite{milt17-96-085007} and \cite{liya16-94-085010} the strong coupling limit is claimed to come out correctly. For a comparison see the recent preprint \cite{milt1808.03816}.
}

At the moment we do not give any interpretation to the negative entropy and restrict ourself to showing its existence within the considered model. We mention that we calculated the complete entropy of the \elm field and the charged fluid. We remind that up to now in Casimir effect configurations negative entropy was observed  only in the interaction part, for instance for two slabs (with dissipation) or two spheres or within the nice model of two oscillators interacting through a third one \cite{hoye16-94-032116}. Thereby only the interaction part of the free energy was considered and with it only the interaction part of the entropy. As a consequence, contributions were left out which could restore an overall positive value of the entropy. \\[.3cm]
{\bf Acknowledgment:} K.K.\ was supported by the Baylor University Summer Sabbatical and Research Leave Program.\\
{The authors thank Kim Milton and his collaborators for the stimulating discussions during the {\it Casimir Effect Workshop June 2018} in Trondheim. }

\end{document}